\documentclass[10pt]{iopart}
\usepackage[T1]{fontenc}
\usepackage[utf8]{inputenc}
\usepackage[english]{babel}
\usepackage{amsmath}
\usepackage[backend=bibtex,style=phys,biblabel=brackets,pageranges=false,url=false,%
maxbibnames=6]{biblatex}
\renewbibmacro{in:}{}
\DeclareFieldFormat[article]{journaltitle}{\textit{#1}}
\AtEveryBibitem{\clearlist{language}}
\preto\fullcite{\AtNextCite{\defcounter{maxnames}{99}}}
\usepackage{physics}
\usepackage{mathtools}
\usepackage{amssymb,bm}
\usepackage[retain-unity-mantissa = false]{siunitx}
\usepackage{slashed}
\usepackage[dvipsnames]{xcolor}
\usepackage{environ}
\usepackage{graphicx}
\usepackage{placeins}
\usepackage{adjustbox}
\usepackage{tabu}
\usepackage{multirow}

\newcommand*{\eh}[1]{\mathrm e^{#1}}
\newcommand*{\NO}[1]{{\bullet}#1\mkern1mu{\bullet}}
\newcommand*{\ext}{\text{ext}}
\NewEnviron{alignt}{%
	\begin{align}
	\begin{aligned}
		\BODY
	\end{aligned}
	\end{align}
}
\makeatletter
\g@addto@macro\bfseries{\boldmath}
\makeatother

\begin{document}
\title[Response of the QED(2) Vacuum]
{Response of the QED(2) Vacuum to a Quench:\\
Long-term Oscillations of the Electric Field and the Pair Creation Rate}
\author{A.~Otto, D.~Graeveling, B.~K\"ampfer}
\address{
Institute of Radiation Physics, Helmholtz-Zentrum Dresden-Rossendorf,\\
01328 Dresden, Germany\\
Institut f\"ur Theoretische Physik, Technische Universit\"at Dresden,\\
01068 Dresden, Germany}
\date{\today}

\begin{abstract}
We consider -- within QED(2) -- the backreaction to the Schwinger pair creation
in a time dependent, spatially homogeneous electric field.
Our focus is the depletion of the external field as a quench and the subsequent
long-term evolution of the resulting electric field.
Our numerical solutions of the self consistent, fully backreacted dynamical
equations exhibit a self-sustaining oscillation of both the electric field and
the pair number depending on the coupling strength.
\end{abstract}

\maketitle
\ioptwocol
\section{Introduction}
The Sauter-Schwinger effect is one of the most important examples of
strong-field QED phenomena.
It refers to the creation of electron-positron pairs by a spatially homogeneous
electric field -- the decay of the vacuum~\cite{sauter,schwinger}
(cf.~\cite{gelis_schwinger_2016} for a recent review).
A common picture is that virtual, entangled pairs constituting the vacuum are
disrupted by the external electric field and lifted on the mass shell, thus
loosing their entanglement in the long-time evolution.
The rate at which pairs are created by an electric field of strength $E$ is
$\propto \exp\qty(-\pi E_c/E)$.
This rate is exceedingly small for macroscopic installations, since the
Sauter-Schwinger (critical) field strength $E_c = m^2/e = \SI{1.3e18}{V/m}$ is
so large, while field strengths presently achievable in the lab are of the order
of $E \approx 0.01E_c$, which results in a huge suppression factor of
$\eh{-300}$.
($m$ and $e$ the electron/positron mass and charge; we employ natural units with
$c = \hbar = 1$.)

Nonetheless one hopes upcoming high-intensity optical laser installations can
provide the avenue towards the necessary fields.
For lasers, the assumption of a constant electric field is not very realistic
and a natural generalization is to let it be time dependent.
This is called the dynamical Schwinger effect.
The special case of a periodic field is dealt with in~\cite{brezin_pair_1970}.
For ideas to boost the pair creation rate by superposing a strong, slowly
varying field with a weak but fast field, see~\cite{schutzhold_dynamically_2008,
dunne_catalysis_2009}.\footnote{
Multi-pair production is considered in~\cite{wollert_multi-pair_2016}.
}
For further generalizations to include spatial gradients,
see~\cite{gies_critical_2016,gies_critical_2017}.
Other setups, not necessarily using lasers, are the field in the vicinity of a
super-heavy atomic nucleus~\cite{greiner_3,rafelski_superheavy_1971,
muller_solution_1972,muller_solution_1973,fillion-gourdeau_enhanced_2013}, or
a superposed XFEL beam~\cite{augustin_nonlinear_2014,di_piazza_effect_2010}.
For a survey of these effects, see~\cite{di_piazza_extremely_2012}.
Furthermore, ideas have also been put forward to forgo the direct detection of
the produced fermions and instead focus on secondary photon
signatures~\cite{karbstein_stimulated_2015,otto_afterglow_2017,
gies_all-optical_2018,blinne_all-optical_2018}.

These investigations have one thing in common:
They suppose the electric (or more generally electro-magnetic) field has no
dynamics of its own and is thus unaffected by the created pairs.
However, physical intuition suggests the electrons and positrons will produce a
current which will in turn generate a counter-acting electro-magnetic field that
gets added to the original one.
One calls this the backreaction of the fermions on the Maxwell field.
Because the Schwinger pair production rate is already strongly suppressed for
$E < E_c$, the study of this further diminishing effect was postponed in favor
of searching for amplification effects.
But since the backreaction is of principal interest in its own right in
illuminating the non-perturbative character of the Schwinger effect, we
reconsider it in this paper in the context of $1+1$~QED (QED(2), also called the
massive Schwinger model~\cite{coleman_more_1976}).
Backreactions were considered within QED(2) in~\cite{chu_capacitor_2010,
mihaila_fermion_2008,mihaila_backreaction_2009}.

A self consistent description of the backreaction is thus needed.
This was first accomplished e.g.\ in~\cite{kluger_pair_1991,kluger_fermion_1992,
kluger_pair_1993,bloch_pair_1999}.
In the present contribution we build on theirs and extend it to investigate the
long term evolution of the electric field and to investigate how the
backreaction affects the created pairs.
Put in a nutshell, we consider the response of the QED(2) vacuum to a quench
caused by an external electric field.

\section{Quantum kinetic equations with backreaction}
We use the framework of the quantum kinetic equations, and our derivation
follows~\cite{schmidt_non-markovian_1999,schmidt_quantum_1998}.
Incorporating the backreaction is done similarly to~\cite{bloch_pair_1999}.
Our starting point is the Dirac equation in $1+1$ dimensions,
$(i\slashed\partial - e\slashed A + m)\Psi = 0$.
The gamma matrices are chosen as $\gamma^0 = \smqty(0 & 1\\ 1 & 0)$ and
$\gamma^1 = \smqty(0 & 1\\ -1 & 0)$.
Since our background field is assumed to be spatially homogeneous, an ansatz for
$\Psi$ via a Fourier transform, $\Psi(t,x) = \int\!\tfrac{\dd{p}}{2\pi}\,
\psi(t,p)\eh{ipx}$ reduces the Dirac equation to a Schrödinger form
\begin{alignt}
& i\dot\psi(t,p) = h(t,p)\psi(t,p),\\
& h(t,p) = \mqty(-p+eA(t) & m\\ m & p-eA(t)).
\end{alignt}
The Hamiltonian $h$ has two time dependent eigenvectors $U$, $V$ to the
eigenvalues $\pm\Omega$ with $\Omega(t,p) = \sqrt{m^2 + \smash{(p-eA(t))}^2}$.
Two linearly independent solutions to the Dirac equations are obtained via the
ansatz
\begin{alignt}
u(t,p) &= \alpha(t,p)U(t,p) + \beta(t,p)V(t,-p),\\
v(t,-p) &= -\beta^*(t,p)U(t,p) + \alpha^*(t,p)V(t,-p)
\label{UV to uv}
\end{alignt}
with $\abs{\alpha}^2 + \abs{\beta}^2 = 1$, resulting in equations for $\alpha$
and $\beta$:
\begin{alignt}
\dot\alpha = -i\Omega\alpha + \frac{eEm}{2\Omega^2}\beta\qc
\dot\beta = -\frac{eEm}{2\Omega^2}\alpha + i\Omega\beta.
\label{alpha beta ODE}
\end{alignt}
To get to observable quantities, we pass over to second quantization by
promoting $\psi$ to an operator on Fock space.
Since we have two bases ($u$, $v$ and $U$, $V$) at our disposal, we can expand
$\psi$ in both:
\begin{alignt}
\psi(t,p) &= c(p)u(t,p) + d^\dagger(-p)v(t,-p)\\
&= C(t,p)U(t,p) + D^\dagger(t,-p)V(t,-p).
\end{alignt}
The relation between $c$, $d$ and $C$, $D$ follows from~\eqref{UV to uv} as
\begin{alignt}
C(t,p) &= \alpha(t,p)c(p) - \beta^*(t,p)d^\dagger(-p),\\
D^\dagger(t,-p) &= \beta(t,p)c(p) + \alpha^*(t,p)d^\dagger(-p),
\end{alignt}
which is called the Bogoliubov transform.
Both sets of operators are fermionic creation/annihilation operators.
The vacuum $\ket0$ is annihilated by $c$, $d$, and the number of produced pairs
is $\expval{C^\dagger C}{0} = \expval{D^\dagger D}{0} = \abs{\beta}^2$.
This lets us define the total pair number
\begin{alignt}
n(t) = \int\!\frac{\dd{p}}{2\pi}\, \abs{\beta(t,p)}^2.
\end{alignt}
The second-quantized Hamiltonian is $H(t) = \int\!\tfrac{\dd{p}}{2\pi}\,
\psi^\dagger(t,p)h(t,p)\psi(t,p)$ and when we normal order it in terms of $C$
and $D$ (indicated by $\NO{\dots}$) we get
\begin{alignt}
\NO{H} = \int\!\frac{\dd{p}}{2\pi}\Omega\qty[C^\dagger C + D^\dagger D].
\end{alignt}
Its expectation value, the energy of the vacuum, is $\expval{\NO{H}}{0}
= V \times 2\int\!\tfrac{\dd{p}}{2\pi} \abs{\beta}^2\, \Omega$.
The factor $V = 2\pi\delta(0)$ is the volume of the system and divergent because
of the homogeneity.

The particles produced by the electric field will induce a current $j^\mu(t,x)
= e\bar\Psi(t,x)\gamma^\mu\Psi(t,x)$.
We denote by its mean field part the normal ordered expectation value
$\bar j^\mu = \expval{\NO{j^\mu}}{0}$ which will be constant in $x$, again due
to the homogeneity.
It evaluates to $\bar j^0 = 0$ and
\begin{alignt}
\bar j^1 = 2e\!\int\!\frac{\dd{p}}{2\pi} \qty[\abs{\beta}^2 \frac{p-eA}{\Omega}
- \Re(\alpha^*\beta)\frac{m}{\Omega}].
\end{alignt}
Note that without normal ordering w.r.t.\ $C$, $D$, $\bar j^0 \ne 0$, which is
unphysical since the external field cannot create a net charge.
Also normal ordering w.r.t.\ $c$, $d$ would yield $\bar j^0 = \bar j^1 = 0$.
This is also unphysical as the particles would not create a (spatial) current.

To this (internal) mean field current we add an arbitrary external current
$j_\text{ext}^0 = 0$, $j_\text{ext}^1 = -\dot E_\ext$ which generates the
external electric field (that is the quench) and plug both into Maxwell's
equation:
\begin{alignt}
\dot E &= \dot E_\ext - 2e\!\int\!\frac{\dd{p}}{2\pi} \qty[\abs{\beta}^2
\frac{p-eA}{\Omega} - \Re(\alpha^*\beta)\frac{m}{\Omega}],\\
\dot A &= -E.
\label{E ODE}
\end{alignt}

An alternative way of arriving at this equation, pursued
in~\cite{bloch_pair_1999}, is to start with the total energy density of the
system
\begin{alignt}
\epsilon = \frac{E^2}{2} + 2\int\frac{\dd{p}}{2\pi}\, \abs{\beta}^2 \Omega
- \int\!\dd{t} \dot E_\ext E
\label{energy conservation}
\end{alignt}
and setting $\dot\epsilon = 0$.
The last term is the work the external current must do to counteract the
electric field.
We use the energy density to check the accuracy of the numerics.

Whichever way one chooses, the equations~\eqref{alpha beta ODE} and~\eqref{E
ODE} together with the initial conditions
\begin{alignt}
&\alpha(t_0, p) = 1\qc \beta(t_0, p) = 0\qc A(t_0) = 0,\\
&E(t_0) = E_\ext(t_0) = 0
\end{alignt}
form a well defined system of coupled ordinary differential equations that we
are going to evaluate.
Note that without the integral incorporating the backreaction in~\eqref{E ODE},
the different momentum modes would be decoupled.

In $1+1$ dimensions, the coupling strength has dimension $[e]=1$.
The fine structure constant is then defined as $\alpha = e^2/4\pi m^2$ and we
give its value when specifying the strength of the backreaction.

\section{Schwinger pair production for various pulse shapes}
We will employ two different quenches caused by external electric fields.
\subsection{Sauter pulse}
The first is the so called Sauter pulse
\begin{alignt}
E_\ext(t) = \frac{E_0}{\cosh^2(t/\tau)}.
\label{Sauter pulse}
\end{alignt}
Note the initial condition $E_\ext(t_0)=0$ can only be approximately fulfilled,
but to arbitrary precision by choosing $-t_0$ sufficiently large.
\begin{figure}
\begin{adjustbox}{center}
\includegraphics{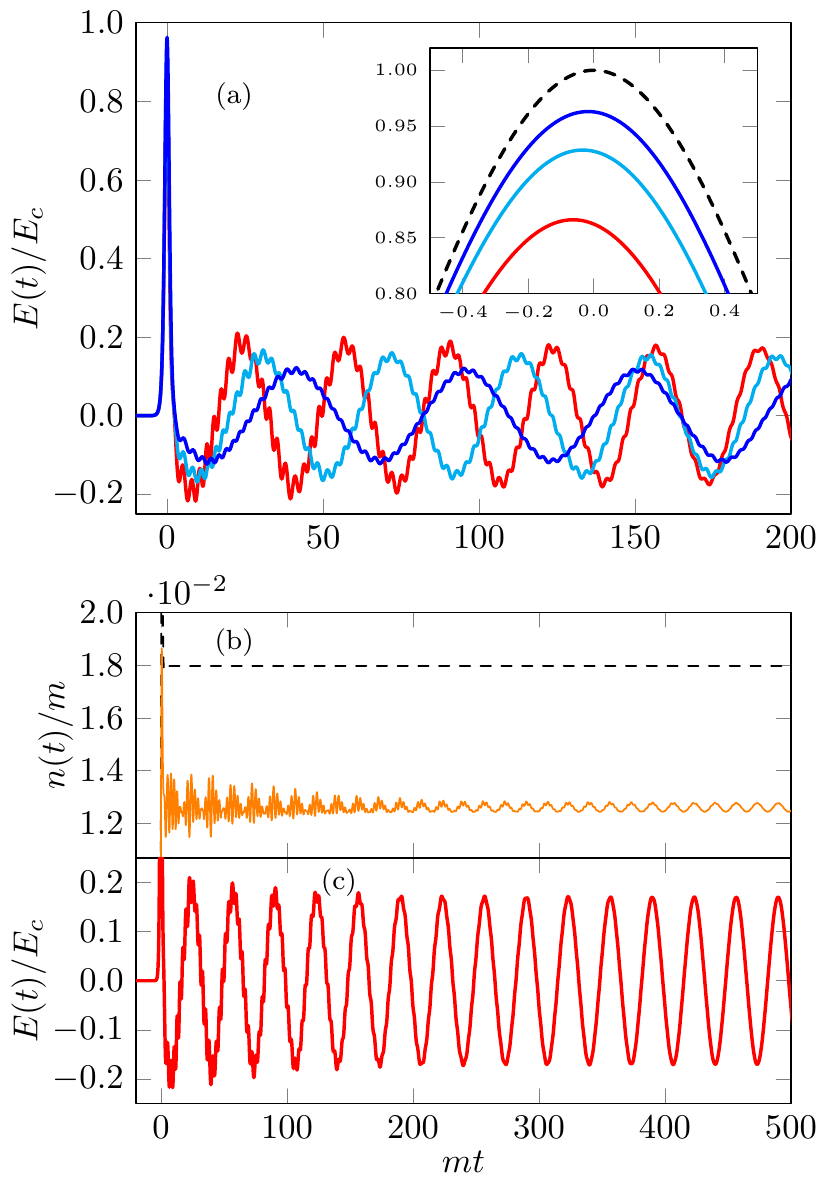}
\end{adjustbox}
\caption{(a) Plot of the electric field as a function of time for the Sauter
pulse~\protect\eqref{Sauter pulse} as external field for $E_0=1E_c$, $\tau=1/m$
and $\alpha=0.05$ (\textcolor{blue}{blue}), $\alpha=0.1$
(\textcolor{cyan}{cyan}), $\alpha=0.2$ (\textcolor{red}{red}).
The inset is a zoom on the peak around $t=0$, where additionally the external
electric field is plotted by the black dashed curve.
(b) The total pair number for $\alpha=0.2$ (\textcolor{orange}{orange}) and
$\alpha=0$, i.e.\ without backreaction (black dashed).
(c) The same as in (a) but only for $\alpha=0.2$ (\textcolor{red}{red}) over a
large time interval.}
\label{fig Sauter}
\end{figure}

In figure~\ref{fig Sauter}(a) we show the time evolution of the electric field,
determined by~\eqref{E ODE}, with the Sauter pulse~\eqref{Sauter pulse} as
external field $E_\ext$.
The first spike is $E$ closely following $E_\ext$.
After the latter has faded away, $E$ starts to settle into a superposition of
oscillations.
These have already been noted in~\cite{kluger_pair_1993,bloch_pair_1999}, and
were also found in~\cite{hebenstreit_simulating_2013} using different methods,
where the Maxwell field was calculated using statistical averages, and
in~\cite{buyens_real-time_2017}, using matrix product states.
In~\cite{petrov_plasmons_2016}, a similar effect was found without a driving
external field, which the authors call plasmons in QED vacuum and attribute to
the vacuum charge polarization.
The inset in figure~\ref{fig Sauter}(a) shows a zoom to the peak of the electric
field around $t=0$.
Increasing the coupling strength screens the electric field more, resulting in
a lower net maximum; i.e.\ the electric field is depleted
(cf.~\cite{seipt_depletion_2017}).
Increasing $\alpha$ also increases both the frequency and amplitude of the
oscillations.

Figure~\ref{fig Sauter}(c) shows the long-term evolution of the electric field
$E$.
For the red curve, one can see that the oscillations with higher frequencies are
transient, and the smaller wiggles vanish.
This is also the case for the other curves, but harder to see.
We conclude that after a short quench by the transient Sauter pulse the system
does not return to its initial state, but rather the electric field keeps
oscillating with a constant frequency and amplitude, looking like an eternal
wobbling.
The pair number (see figure~\ref{fig Sauter}(b)) oscillates in a similar manner
as the electric field.
Both subsystems are coupled through the total energy conservation~\eqref{energy
conservation}.
Switching off the backreaction, i.e.\ ignoring the integral in~\eqref{E ODE},
yields no oscillations after the external field declines to zero, with both $E$
and $n$ constant.

\subsection{Flat-top $C^\infty$ pulse}
The second pulse shape we employ is a $C^\infty$ pulse with the following
properties:
\begin{alignt}
E_\ext(t) = \begin{cases}
0, & t \le 0,\\
E_0, & t_r \le t \le t_r+t_f,\\
0, & t \ge 2t_r+t_f,
\end{cases}
\label{const bump pulse}
\end{alignt}
and monotonously increasing/decreasing where not specified.
Its precise construction can be found in the appendix.
In contrast to the Sauter pulse, it has two time scales, the ramping time $t_r$
over which the electric field is switched on and off, and the flat top time
$t_f$ over which it is constant.
The case $t_f \to \infty$ captures the plain\footnote{
Rather a modified version, where the electric field does not extend to the
infinite past, but gets turned on smoothly at some time.
}
Schwinger effect, but with the additional backreaction.
The backreaction again causes some depletion, as evidenced in figure~\ref{fig
const bump}(a):
\begin{figure}
\begin{adjustbox}{center}
\includegraphics{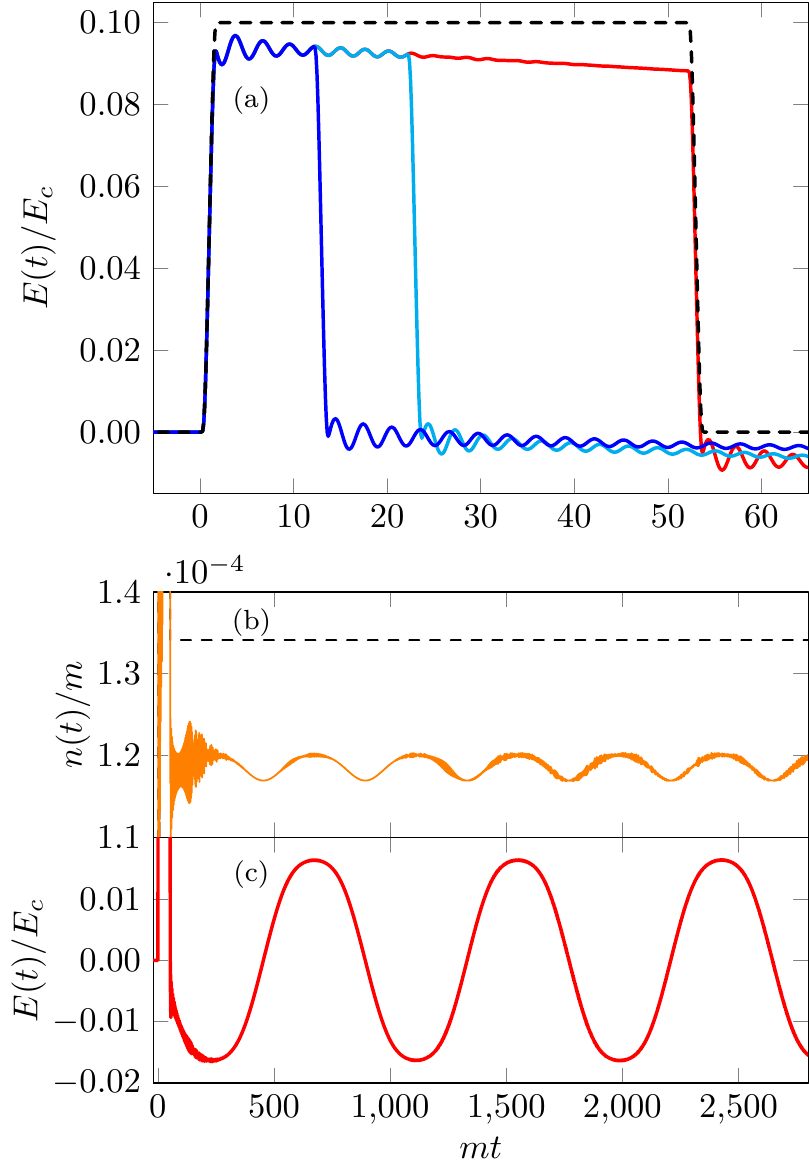}
\end{adjustbox}
\caption{(a) Plot of the electric field as a function of time for the
$C^\infty$ pulse~\protect\eqref{const bump pulse} as external field for
$E_0=0.1E_c$, $t_r=2/m$, $\alpha=0.1$ and $t_f=10/m$ (\textcolor{blue}{blue}),
$t_f=20/m$ (\textcolor{cyan}{cyan}), $t_f=50/m$ (\textcolor{red}{red}).
The external electric field is plotted in black dashed for $t_f=50/m$.
(b) The total pair number for $t_f=50/m$ (\textcolor{orange}{orange}) and
without backreaction (black dashed1).
(c) The same as in (a) but only for $t_f=50/m$ (\textcolor{red}{red}) over a
large time interval.}
\label{fig const bump}
\end{figure}
The electric fields grows to almost the value of the external field, enters
some transient oscillations, and then drops, synchronized with the drop of the
external field.
Subsequently the field seemingly displays a similar eternal wobbling as in the
long-time regime of the Sauter quench, see figure~\ref{fig const bump}(c).
The pair number's oscillations (figure~\ref{fig const bump}(b)) are also
synchronized with the field oscillations.
\begin{figure}
\begin{adjustbox}{center}
\includegraphics{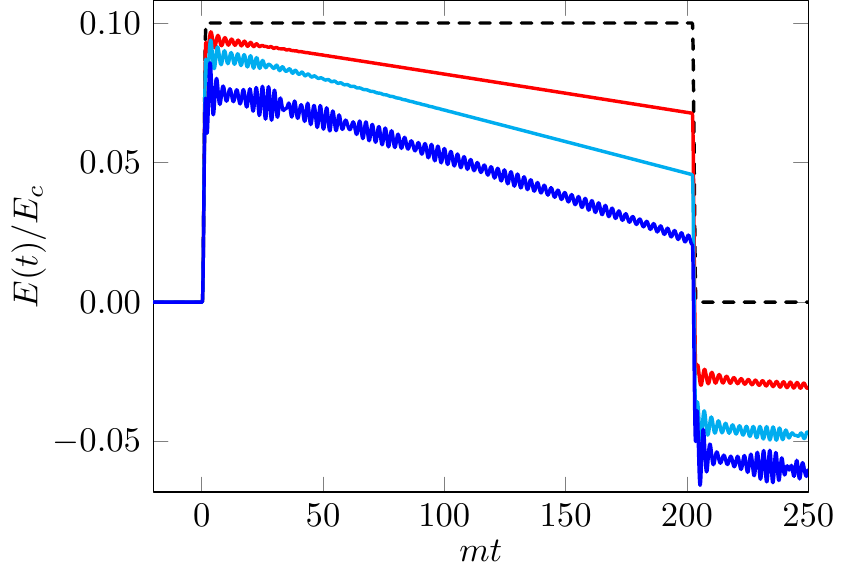}
\end{adjustbox}
\caption{Plot of the electric field as a function of time for the $C^\infty$
pulse~\protect\eqref{const bump pulse} as external field for $E_0=0.1E_c$,
$t_r=2/m$, $t_f=200/m$ and $\alpha=0.1$ (\textcolor{red}{red}), $\alpha=0.2$
(\textcolor{cyan}{cyan}), $\alpha=0.5$ (\textcolor{blue}{blue}).
The external electric field is plotted in black dashed.}
\label{fig const bump alpha}
\end{figure}

The depletion can be more clearly seen in figure~\ref{fig const bump alpha},
where we plot the electric field for a longer lasting $C^\infty$ pulse and for
varying $\alpha$.
The sudden change in the external field lifts the total field but also induces
transient oscillations.
Once the external field is constant, the total field starts declining with a
slope that grows as $\alpha$ does, while the oscillations tend to zero (but
observe that they take longer to do so for larger coupling strengths).
The switching-off of the external field makes the total field swing in the
oppsite direction after which it enters the long-term oscillating state
exhibited in figure~\ref{fig const bump}(c).
Note that per~\eqref{energy conservation} the external field does not add energy
to the system for $E_\ext=\text{const}$.
Thus in the flat-top section the energy just gets shifted from the electric
field to the created fermions.

\section{Summary}
In summary we consider the impact of the backreaction on Schwinger type pair
creation.
The produced pairs screen the external field and facilitate its net depletion.
Viewing the external field as a quench to the vacuum, it is interesting to see
the vacuum response as a wobbling of the number of created pairs in phase with
the long-term oscillations of the induced electric field.
For the selected examples and within the considered time intervals, the vacuum
looks like an eternally swinging medium.
Due to numeric reasons (see the momentum integral in~\eqref{E ODE}) we worked
in $1+1$ dimensional QED, i.e.\ QED(2).
The swinging vacuum response, however, seems to be generic, as the examples
in~\cite{kluger_pair_1993,bloch_pair_1999,hebenstreit_simulating_2013,
buyens_real-time_2017,petrov_plasmons_2016} show.

\bigskip\noindent\textbf{Acknowledgments:}
The authors gratefully acknowledge inspiring discussions with R.~Schützhold,
H.~Gies, R.~Alkofer, D.~B.~Blaschke and C.~Greiner.
Many thanks go to S.~Smolyansky and A.~Panferov for previous common work on the
plain Schwinger process.
The fruitful collaboration with R.~Sauerbrey and T.~E.~Cowan within the HIBEF
project promoted the present investigation.

\appendix
\section{Construction of the $C^\infty$ pulse}
To construct the pulse shape for the electric field~\eqref{const bump pulse} we
use a procedure often employed in differential geometry for partitions of
unity, see e.g.\ chapter~13 in~\cite{tu_introduction_2008}.

First define
\begin{alignt}
r(x) = \begin{cases}
0, & x \le 0,\\
\eh{-\frac 1x}, & x > 0.
\end{cases}
\end{alignt}
This function is $C^\infty$ but not analytic, since $r^{(n)}(0) = 0$.
Using it, define $s(x) = r(x)/[r(x)+r(1-x)]$.
Observe $s(x\le0) = 0$ and $s(x\ge1) = 1$.
This lets us define
\begin{alignt}
E_\ext(t) = E_0s\qty(\frac{t}{t_r})s\qty(\frac{2t_r+t_f-t}{t_r})
\end{alignt}
which has all the properties we claimed for $E_\ext$ in~\eqref{const bump
pulse}.
\printbibliography
\end{document}